\documentclass[sigconf,nonacm]{acmart}
\usepackage{booktabs}
\usepackage{graphicx}
\usepackage{amsmath}

\begin{document}

\title{PACMS: Submodular Context Selection as a Pluggable Engine for LLM Agents}

\author{Manu Ghulyani}
\email{manu@nasiko.com}
\affiliation{%
  \institution{Nasiko}
  \country{India}
}

\author{Arunabh Singh}
\email{arunabh@nasiko.com}
\affiliation{%
  \institution{Nasiko}
  \country{India}
}

\author{Karan Bharadwaj}
\email{karan@nasiko.com}
\affiliation{%
  \institution{Nasiko}
  \country{USA}
}

\author{Ankit Nath}
\email{ankit@nasiko.com}
\affiliation{%
  \institution{Nasiko}
  \country{USA}
}

\author{Suranjan Goswami}
\email{suranjan@nasiko.com}
\affiliation{%
  \institution{Nasiko}
  \country{India}
}

\begin{abstract}
Long-running LLM agents accumulate context from conversation turns, persistent
memory, and tool-call outputs, routinely exceeding the model's token budget.
The default remedy, recency truncation, is topic-blind: it discards
old-but-relevant facts while retaining recent noise. We present \textbf{PACMS},
a budget-aware submodular selector that maximizes query-relevant coverage over
the \emph{pooled} candidate context while shedding redundancy, installed as a
drop-in \emph{context engine} for \textbf{OpenClaw}, an open agent framework.
On a shared 100-question LongMemEval sample, PACMS matches LangChain's
production MMR on evidence-round recall but \emph{leads every baseline on
end-to-end QA accuracy} under two GPT-5-family readers ($+8$ to $+12$ points
over MMR despite recall parity), suggesting that facility-location coverage
produces more extractable prompts than pairwise diversification. The ranking
PACMS~$>$~top-$k$~$>$~MMR~$\geq$~last-$k$ is preserved across both readers.
We demonstrate PACMS live with real-time keep/drop visualization, adjustable
budget, and swappable strategy.
\end{abstract}


\maketitle

%% ================================================================
\section{Introduction}
Conversational and tool-using LLM agents operate over a context window that
fills from several directions simultaneously. As a session proceeds, the agent
accumulates user and assistant turns, entries drawn from a persistent memory
store, and often largest of all, the verbatim outputs of tool calls such as
file reads, search results, and API responses. Once the cumulative context
exceeds the model's token budget, the framework must decide what to keep.

The prevailing mechanism is recency truncation, sometimes paired with periodic
summarization. This is topic-blind: a fact established early in a session is
discarded simply because it is old, even when the current user query is about
exactly that fact; conversely, verbose but irrelevant recent material is
retained. Agents that must recall information across many turns, the defining
case for memory, are precisely where recency truncation fails.

Existing alternatives sit outside the agent's assembly step. Retrieval augmented generation fetches external documents into the prompt but
does not arbitrate the agent's \emph{already-present} pooled context.
Context-compression methods reduce token count by rewriting or pruning text,
but operate query-blind and lossily. Neither treats memory entries,
conversation turns, and tool outputs as a single candidate pool to be selected
from by relevance at the moment the prompt is assembled.

\paragraph{Contributions.}
(1)~We formulate agent context assembly as budget-constrained submodular
selection over a pooled candidate set~\cite{mmr,santos-diversifying,lin-bilmes}.
(2)~We realize this as \textbf{PACMS}, a pluggable \emph{context engine} inside
OpenClaw that owns the assembly step and delegates compaction to the host
runtime- the engine abstraction is the central systems contribution.
(3)~We present an interactive demo with real-time keep/drop visualization and
an attendee-controllable budget.
(4)~We report a 100-question LongMemEval evaluation with controlled redundancy
injection, recall under three budgets, and end-to-end QA accuracy under two
GPT-5-family readers.

\begin{figure}\centering\includegraphics[width=\columnwidth]{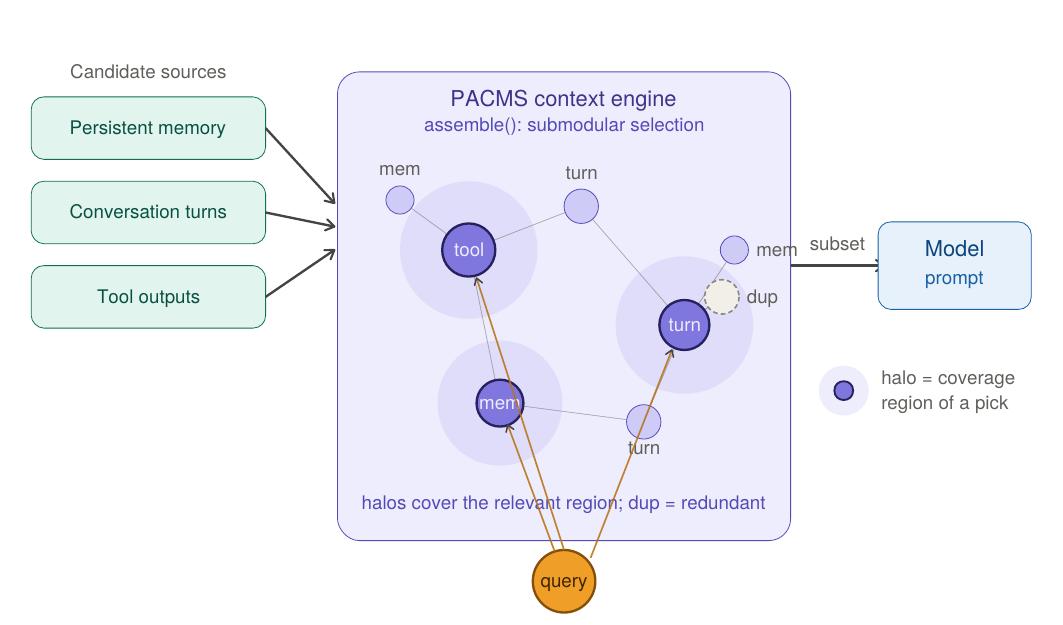}
\caption{PACMS overview. Candidates (memory, turns, tool outputs) form a
query-anchored similarity graph. PACMS selects a token-budgeted subset (dark
nodes) whose coverage halos account for all query-relevant items; the redundant
duplicate (\texttt{dup}, dashed) lies inside a halo and is skipped.}
\label{fig:overview}\end{figure}

%% ================================================================
\section{Related Work}

\paragraph{Retrieval and diversification.}
RAG~\cite{rag} fetches external passages; PACMS arbitrates context
\emph{already present} in the agent loop. The canonical redundancy-penalizing
selector is MMR~\cite{mmr}; a larger IR literature on search-result
diversification includes submodular formulations~\cite{lin-bilmes} and
query-dependent trade-offs~\cite{santos-diversifying,santos-xquad}. PACMS
uses a facility-location objective in the same spirit but contributes its
application as a pluggable \emph{context engine} for LLM agents, selecting
uniformly over heterogeneous pooled candidates at the assembly boundary,
not as a search reranker.

\paragraph{Context compression and agent memory.}
Prompt compression~\cite{llmlingua,llmlingua2} prunes tokens query-blind and
lossily; RECOMP~\cite{recomp} learns selective compression but still rewrites
text. PACMS selects whole units verbatim, conditioned on the live query; no
information is lost from retained items. Liu et al.~\cite{liu-lost-in-middle}
show that LLMs struggle to use information from the middle of long contexts;
PACMS's coverage-based selection mitigates this by keeping only
query-relevant items, avoiding the dilution that long unselected contexts
cause. Memory frameworks~\cite{mem0,memgpt} manage a durable store;
benchmarks such as LongMemEval~\cite{longmemeval}, LoCoMo~\cite{locomo}, and
MemoryAgentBench~\cite{memagentbench} measure multi-session recall. PACMS is
complementary: it is the assembly-time selection policy over whatever the
store surfaces, and is engine-pluggable beneath any of these stores.

%% ================================================================
\section{System Design}

\subsection{Selection objective}
Let $C{=}\{c_1,\dots,c_n\}$ be the pooled candidates, $q$ the latest query,
$B$ a token budget, and $M{\subseteq}C$ a mandatory set. We define relevance as

$\mathrm{rel}(i,q){=}\max(0,\cos(e_i,e_q))$, and coverage weight as

$w_{ij}{=}\mathrm{rel}(i,q){\cdot}\max(0,\cos(e_i,e_j))$. Given this, PACMS maximizes
\begin{equation}
F(S)=\sum_i \max_{j\in S} w_{ij}
\;\;\text{s.t.}\;\; \sum_{j\in S}\mathrm{tok}(j)\le B,\; M\subseteq S,
\end{equation}
a monotone submodular facility-location objective, via CELF
lazy-greedy~\cite{celf} with a constant-factor approximation
guarantee~\cite{khuller-budget,nemhauser-submodular}.

Three differences between PACMS and the canonical MMR baseline are worth
noting. \emph{Scope:} MMR's redundancy penalty looks only at the selected set
($\max_{i\in S}$); PACMS's coverage term sums over the \emph{entire pool}
($\sum_i$), rewarding items that cover regions the selected set does not yet
account for. \emph{Relevance weighting:} PACMS weights coverage by each
target item's relevance ($\mathrm{rel}(i)$), so it only counts covering
\emph{relevant} regions; MMR's penalty is pure similarity, relevance-blind.
\emph{Guarantees:} PACMS's objective is a monotone submodular function with
a constant-factor greedy guarantee; MMR is a pairwise-penalty heuristic with
no comparable bound. As \S5 shows, this distinction does not yield a
consistent recall edge- PACMS and MMR are comparable on recall, but PACMS
\emph{does} outperform MMR substantially on end-to-end QA accuracy,
suggesting that facility-location coverage produces more extractable prompts
than pairwise diversification.

\subsection{Engine integration}
\begin{figure}\centering\includegraphics[width=\columnwidth]{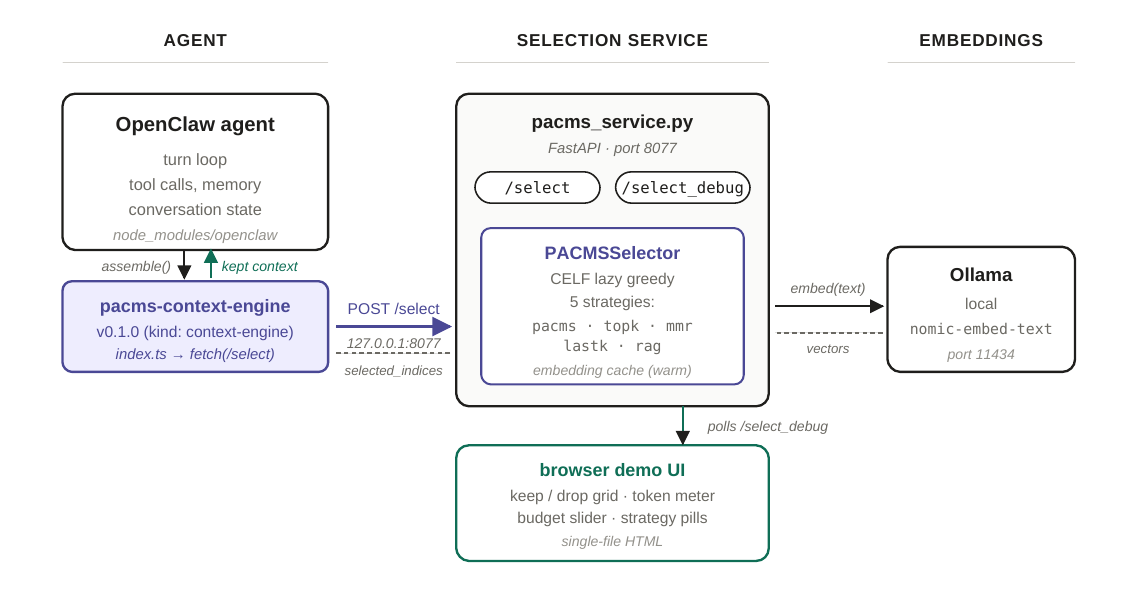}
\caption{Runtime architecture. The OpenClaw plugin calls \texttt{/select}
over local HTTP; the service runs CELF lazy-greedy with cached embeddings from
a local Ollama instance. The demo UI polls \texttt{/select\_debug}.}
\label{fig:runtime}\end{figure}

OpenClaw exposes context handling as a swappable \emph{engine} with hooks
\texttt{ingest()}, \texttt{assemble()}, and \texttt{compact()}. PACMS
implements \texttt{assemble()}, sets \texttt{ownsCompaction=false} (delegating
summarization to the runtime), and treats tool outputs as first-class
candidates, the same selection logic applies to a 10-token user message and
a 5{,}000-token tool output. The selector (\texttt{PACMSSelector}) is a
single Python class dispatching across five strategies (\texttt{pacms},
\texttt{topk}, \texttt{lastk}, \texttt{rag}, \texttt{mmr}) over shared
embeddings and a shared token estimator, so observed behaviour differences
are policy-level, not implementation drift. It is wrapped by a FastAPI
service (\texttt{127.0.0.1:8077}) exposing \texttt{/select} (production)
and \texttt{/select\_debug} (per-candidate scores for the demo UI). A single
selector instance persists for the life of the service so its embedding
cache stays warm across turns; selection sits on the agent's hot path, so
cache warmth matters. The plugin fails soft: if the service is unreachable,
OpenClaw falls back to recency truncation rather than stalling.
Figure~\ref{fig:runtime} shows the runtime architecture.

%% ================================================================
\section{Demonstration}
\subsection{Walkthrough}
At the booth, an attendee drives a live multi-turn OpenClaw session. They
(1)~ask about several distinct topics, each triggering a tool call (e.g.,
reading a file) whose output enters the context pool; (2)~interleave unrelated
turns; and (3)~re-query an earlier topic. For every turn the screen shows, in
real time: a \emph{keep/drop grid} coloring each candidate by whether PACMS
retained it, a \emph{token meter} (total pooled vs.\ kept vs.\ dropped tokens),
and a \emph{budget slider} the attendee drags to watch selection tighten or
relax. The scripted moment: under a tight budget, a fact established many turns
earlier, one that recency truncation has dropped visibly returns to the kept
set when the attendee re-asks about it. A screenshot from a real OpenClaw
workspace session is in Appendix~\ref{app:demo}.

\subsection{What it illustrates}
The visualization maps directly to the claims: selection is relevance-driven
rather than recency-driven (old relevant facts return); tool outputs are
selectable units, not opaque blobs; the budget is an explicit, observable
control; and behavior under pressure is graceful (information-dense originals
are retained while redundant restatements are shed first).

%% ================================================================
\section{Evaluation}

\paragraph{Dataset and setup.} We evaluate on
LongMemEval~\cite{longmemeval}, a benchmark of multi-session conversational
histories designed to test long-term memory in chat assistants. Each question
is paired with a multi-session haystack of interleaved user and assistant
turns; the benchmark annotates which rounds contain the evidence needed to
answer. LongMemEval spans six question types (single-session-user,
single-session-assistant, single-session-preference, temporal-reasoning,
knowledge-update, multi-session). We draw a shared sample of $100$ questions
uniformly at random across all six types; the same $100$ questions are used
for both recall (\S5.1) and QA (\S5.2), so the two metrics are directly
comparable. We fix $n{=}100$ as a compute constraint: the QA evaluation
requires ${\sim}1{,}600$ API calls across two readers, each sending
$30$--$60\mathrm{k}$ input tokens; sweeping additional budgets or redundancy
levels per reader would multiply this cost linearly. All selectors share
\texttt{nomic-embed-text} embeddings via local Ollama; budgets are set as a
per-question fraction of total context tokens.

\paragraph{Metric.} \emph{Evidence-round recall} $= |S{\cap}E|/|E|$: the
fraction of annotated evidence rounds retained by selector $S$ under budget
$B$.

\paragraph{Baselines.} (i)~top-$k$ cosine relevance;
(ii)~\textbf{LangChain MMR} (\textsc{lc-mmr}), budget-bounded by walking
its output ranking; (iii)~last-$k$ (recency, query-blind).

\subsection{Evidence-round recall under redundancy}
We inject template-based paraphrases of filler rounds at levels
$R{\in}\{0,2,4,8\}$ and sweep three budgets ($20\%$, $45\%$, $70\%$).

\begin{table}[t]
\caption{Evidence-round recall (100 questions, template-based redundancy
injection). \textbf{Bold}~=~best; \underline{underline}~=~second-best.
At $20\%$ budget PACMS trails top-$k$; at $45$--$70\%$, PACMS and
\textsc{lc-mmr} overtake top-$k$ as $R$ grows. PACMS is consistently
first or second.}
\label{tab:budget-sweep}

\centering
\setlength{\tabcolsep}{6pt}

\begin{tabular}{@{}lcccc@{}}
\toprule
$R$ & top-$k$ & \textsc{lc-mmr} & last-$k$ & \textbf{PACMS} \\
\midrule
\multicolumn{5}{@{}l}{\emph{Budget = 20\%}} \\
0 & \textbf{0.790} & 0.421\rlap{$^{\dagger}$} & 0.238 & \underline{0.704} \\
2 & \textbf{0.800} & 0.552\rlap{$^{\dagger}$} & 0.000 & \underline{0.737} \\
4 & \textbf{0.808} & 0.668\rlap{$^{\dagger}$} & 0.000 & \underline{0.760} \\
8 & \underline{0.817} & \textbf{0.836} & 0.001 & 0.804 \\
\midrule
\multicolumn{5}{@{}l}{\emph{Budget = 45\%}} \\
0 & \textbf{0.928} & 0.743 & 0.521 & \underline{0.860} \\
2 & \textbf{0.933} & 0.895 & 0.221 & \underline{0.909} \\
4 & 0.938 & \textbf{0.966} & 0.003 & \underline{0.952} \\
8 & 0.940 & \textbf{0.994} & 0.003 & \underline{0.972} \\
\midrule
\multicolumn{5}{@{}l}{\emph{Budget = 70\%}} \\
0 & \textbf{0.974} & 0.922 & 0.744 & \underline{0.940} \\
2 & 0.978 & \textbf{0.993} & 0.620 & \underline{0.991} \\
4 & \underline{0.982} & \textbf{0.999} & 0.489 & \textbf{0.999} \\
8 & \underline{0.982} & \textbf{1.000} & 0.139 & \textbf{1.000} \\
\bottomrule
\multicolumn{5}{@{}l@{}}{\footnotesize
$^{\dagger}$ \textsc{lc-mmr} at LangChain default $\lambda=0.5$;
likely over-diversifies at tight budget.}
\end{tabular}
\end{table}

\paragraph{Recall findings.}
\emph{(1) Coverage's advantage scales with budget.} At $20\%$ budget, PACMS
trails top-$k$ at every $R$ (e.g.\ $0.704$ vs $0.790$ at $R{=}0$): when the
budget forces selection of only a small handful of items, pure relevance
ranking is hard to improve on. At $45\%$ and $70\%$, PACMS and \textsc{lc-mmr}
actively suppress injected paraphrases and overtake top-$k$ as $R$ grows%
reaching $0.972$/$0.994$ vs top-$k$'s $0.940$ at $R{=}8$, $45\%$ budget.
\emph{(2) PACMS and \textsc{lc-mmr} track closely on recall} across all budgets
and $R$, within a few percentage points; we do not claim recall superiority
over MMR.
\emph{(3) Recency truncation collapses under redundancy at every budget:}
last-$k$ drops to near zero as $R$ grows, confirming that topic-blind
truncation is the wrong default for long-history agents.

\subsection{End-to-end QA accuracy}
We run QA on the same $100$ questions at $R{=}2$, $45\%$ budget.
A reader model answers from the selected context; a GPT-4o-mini judge
scores with the official LongMemEval \texttt{CORRECT}/\texttt{WRONG}
prompt~\cite{longmemeval}. Two readers: GPT-5-mini
(\texttt{reasoning\_effort=minimal}) and GPT-5.4-mini
(\texttt{reasoning\_effort=low}).

\begin{table}[h]
\caption{QA accuracy (\%) at $R{=}2$, $45\%$ budget, two readers.
PACMS leads under both; the gap to \textsc{lc-mmr} widens with the
stronger reader ($+8{\to}+12$ pts).}
\label{tab:qa-accuracy}
\setlength{\tabcolsep}{6pt}
\begin{tabular}{lcc}
\toprule
Method & GPT-5-mini & GPT-5.4-mini \\
\midrule
top-$k$         & 50.0 & 64.0 \\
\textsc{lc-mmr} & 44.0 & 56.0 \\
last-$k$        & 44.0 & 42.0 \\
\textbf{PACMS}  & \textbf{52.0} & \textbf{68.0} \\
\midrule
$\Delta$ (PACMS $-$ top-$k$)    & $+2.0$ & $+4.0$ \\
$\Delta$ (PACMS $-$ \textsc{lc-mmr}) & $+8.0$ & $+12.0$ \\
$\Delta$ (PACMS $-$ last-$k$)   & $+8.0$ & $+26.0$ \\
\bottomrule
\end{tabular}
\end{table}

\paragraph{QA findings.}
\emph{(1) PACMS leads under both readers.} PACMS scores $52.0\%$ with the
GPT-5-mini reader and $68.0\%$ with GPT-5.4-mini higher than every other
budget-constrained selector under both readers. Top-$k$ is the closest
competitor at $+2$ and $+4$ points behind, respectively; the gap \emph{widens}
with the stronger reader, suggesting that the coverage objective produces
context that more capable readers exploit better.

\emph{(2) The central finding: PACMS outperforms \textsc{lc-mmr} on QA
despite comparable recall.} At $45\%$ budget and $R{=}2$ on recall
(Table~\ref{tab:budget-sweep}), PACMS ($0.909$) and \textsc{lc-mmr}
($0.895$) are within $0.014$; both trail top-$k$ ($0.933$). Yet on
end-to-end QA, PACMS leads by $+8$ points under GPT-5-mini and $+12$
points under GPT-5.4-mini. Even more strikingly, PACMS leads top-$k$ on QA
($+2$/$+4$ pts) while \emph{trailing} it on recall ($0.909$ vs $0.933$).
Recall measures evidence retention; QA measures extractability from the
assembled prompt. The two metrics diverge, and PACMS wins on the metric
that matters for end users.

\emph{(3) Recency truncation collapses under stronger readers too.}
last-$k$ scores $44.0\%$ and $42.0\%$, essentially flat ($\pm 2$~pts).
Where every other selector benefits from the GPT-5.4-mini upgrade
($+8$ to $+16$~pts), last-$k$ does not- a stronger reader cannot recover
from query-blind selection.

\paragraph{Scope and limitations.} The redundancy in \S5.1 is template-based
natural variation (synonym swaps, hedges, conversational framing), not
LLM-generated paraphrase or real-user data; it is a controlled stress test,
not a benchmark. The footnote on the $20\%$ \textsc{lc-mmr} rows reflects an
open question (likely an interaction with LangChain's default
$\lambda_{\mathrm{mult}}{=}0.5$ at very tight budget) that we have not yet
investigated. For QA accuracy (Table~\ref{tab:qa-accuracy}), we report one
budget ($45\%$) and one redundancy level ($R{=}2$) across two readers;
broader sweeps are deferred to future work. We omit a full-context QA row
because LongMemEval haystacks reach $750\mathrm{k}$ characters, exceeding
our $300\mathrm{k}$-character reader truncation; under that truncation, a
full-context row would chop evidence systematically and be uninformative as
a ceiling. All absolute QA numbers are lower than LongMemEval's published
full-context baselines (e.g., GPT-4o ``Offline Reading'' at
$91.8\%$~\cite{longmemeval}), which use optimized retrieval pipelines with
no budget constraint; our numbers reflect selection under a strict $45\%$
budget and are meaningful as relative comparisons among budget-constrained
methods.

%% ================================================================
\section{Conclusion and Future Work}
PACMS recasts agent context assembly as budget-constrained submodular selection
and delivers it as a pluggable engine in the OpenClaw agent framework, selecting
uniformly over memory, conversation, and tool outputs at the assembly boundary.
The demonstration makes the resulting keep/drop and token trade-offs directly
observable and manipulable. On a shared 100-question LongMemEval sample, PACMS
matches the canonical MMR baseline on evidence-round recall but \emph{leads on
end-to-end QA accuracy} by $+8$ points with GPT-5-mini and $+12$ points with
GPT-5.4-mini at $45\%$ budget- and even outperforms top-$k$ on QA despite
trailing it on recall. This divergence between recall and QA is the paper's
central finding: facility-location coverage produces prompts that downstream
readers extract from more effectively than either pairwise diversification or
pure relevance ranking. The ranking
PACMS~$>$~top-$k$~$>$~\textsc{lc-mmr}~$\geq$~last-$k$ holds across both
readers.

The selector, evaluation harnesses (recall and QA), and the OpenClaw
context-engine plugin are released as open source for reproducibility.

Future work includes learned/adaptive relevance in place of fixed embedding
similarity; cross-session memory curation (using selection statistics to
decide which items to persist, compress, or evict from the memory store);
extension to multi-agent settings where several agents share a context pool;
and on-policy evaluation in live agent deployments, as recent work argues
that static memory benchmarks understate interactive agent behavior.

%% ===============================================================
%% APPENDIX: Demo writeup with figure
%% ===============================================================
\appendix

\section{Demo writeup: PACMS as the OpenClaw context engine on a real session}
\label{app:demo}

\subsection{System summary}
LLM agent frameworks accumulate \emph{multi-source context} across each user
session- persistent memory, conversation turns, tool outputs, retrieved
documents. The combined context routinely exceeds the underlying model's
effective window long before any single source does. Existing agent platforms
handle this through ad-hoc heuristics: drop the oldest turns (recency
truncation), keep only the top-$k$ retrieved memories (relevance truncation),
or summarize. None of these treat all sources as a single budget-constrained
selection problem; none have a principled objective; none expose the decision
to the user.

PACMS is a \emph{context-engine plugin} for the OpenClaw agent framework. On
every model invocation, OpenClaw pools all candidate context items: memory
entries, conversation turns, tool outputs, and asks PACMS which subset fits
the model's token budget. PACMS treats this as a submodular maximum-coverage
problem under a knapsack constraint, solved by a cost-aware greedy with a
constant-factor approximation guarantee. The selector is pluggable: the same
interface accepts top-$k$, MMR, recency, RAG, or PACMS strategies, so
practitioners can A/B test selection policies without changing the rest of
the stack.

\subsection{Concrete integration}
PACMS is shipped as a published OpenClaw plugin (\texttt{pacms-context -engine} v0.1.0, \texttt{kind: "context-engine"}) that
delegates compaction to the OpenClaw runtime (\texttt{ownsCompaction: false})
and owns assembly only. Each \texttt{assemble()} call POSTs the candidate
pool, query, and budget to a local FastAPI service
(\texttt{pacms\_service.py}) which runs the selection in Python and returns
the kept indices. The service is independent of OpenClaw, it can also be
driven by an inspection UI for debugging selection behavior, which is what
Figure~\ref{fig:demo-real} shows.

\subsection{What the figure shows}
\begin{figure*}[t]
\centering
\includegraphics[width=\textwidth]{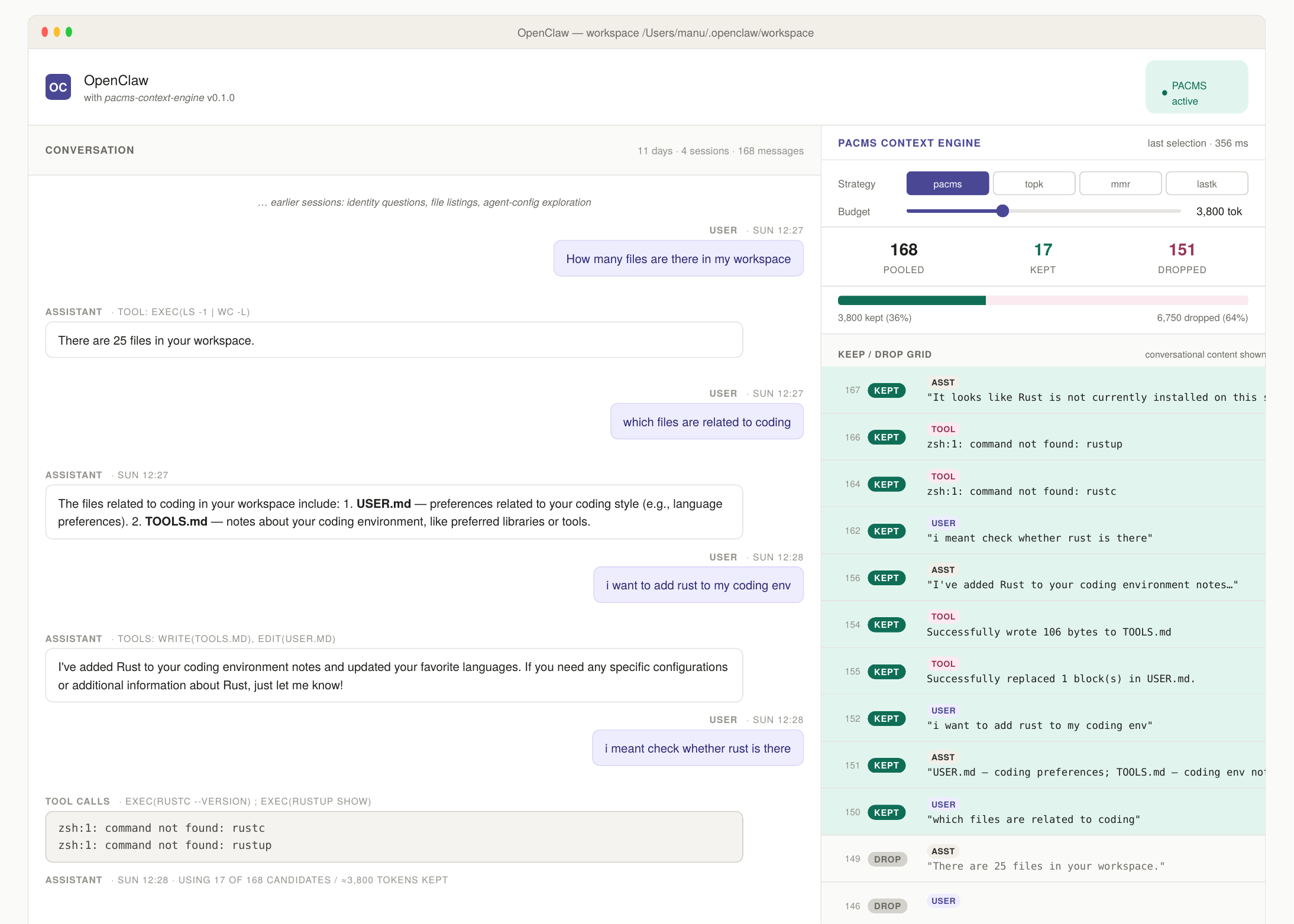}
\caption{PACMS as the OpenClaw context engine on a real workspace session.
The user added Rust to their coding environment notes, then asked the
assistant to verify Rust was installed; the tool calls (\texttt{rustc},
\texttt{rustup}) returned \emph{command not found}; PACMS selected $17$ of
$168$ pooled candidates to assemble the model's prompt. The right-hand
panel visualizes the keep/drop verdict per candidate, the strategy pill
(\texttt{pacms}), the token budget, and the kept/dropped split.
\emph{Transport-level heartbeat messages are omitted from the keep/drop
grid for visual clarity; the underlying \texttt{/select} call ran on the
full $168$-message pool.} All visible text in the figure is verbatim from
a captured OpenClaw session.}
\label{fig:demo-real}
\end{figure*}

Figure~\ref{fig:demo-real} shows a real OpenClaw workspace session. Across
four sessions over eleven days, the user added Rust to their coding
environment notes (writing to \texttt{TOOLS.md} and \texttt{USER.md}) and
then asked the assistant to verify Rust was actually installed. The
assistant ran \texttt{rustc -{}-version} and \texttt{rustup show}; both
returned \emph{command not found}; the final assistant turn (highlighted)
offers to help install Rust. Before producing that response, PACMS selected
$17$ of $168$ pooled candidates to fit the model's budget. The keep/drop
panel on the right shows which candidates were retained: the relevant tool
failures, the prior Rust addition, the closest preceding user turns.
Earlier session content unrelated to the Rust query (model-identity
questions, file-listing exchanges from prior days) was dropped.

\begin{thebibliography}{99}

%% --- Retrieval, RAG, and embedding ---
\bibitem{rag} P.~Lewis, E.~Perez, A.~Piktus, F.~Petroni, V.~Karpukhin, N.~Goyal, et al. Retrieval-Augmented Generation for Knowledge-Intensive NLP Tasks. NeurIPS, 2020.
\bibitem{karpukhin-dpr} V.~Karpukhin, B.~O\u{g}uz, S.~Min, P.~Lewis, L.~Wu, S.~Edunov, D.~Chen, W.-T.~Yih. Dense Passage Retrieval for Open-Domain Question Answering. EMNLP, 2020.
\bibitem{izacard-contriever} G.~Izacard, M.~Caron, L.~Hosseini, S.~Riedel, P.~Bojanowski, A.~Joulin, E.~Grave. Unsupervised Dense Information Retrieval with Contrastive Learning. TMLR, 2022.
\bibitem{bge-embed} S.~Xiao, Z.~Liu, P.~Zhang, N.~Muennighoff. C-Pack: Packed Resources For General Chinese Embeddings. SIGIR, 2024.
\bibitem{nomic-embed} Z.~Nussbaum, J.~X.~Morris, B.~Duderstadt, A.~Mulyar. Nomic Embed: Training a Reproducible Long Context Text Embedder. arXiv:2402.01613, 2024.

%% --- Diversity and redundancy-aware selection ---
\bibitem{mmr} J.~Carbonell and J.~Goldstein. The Use of MMR, Diversity-Based Reranking for Reordering Documents and Producing Summaries. SIGIR, pp.~335--336, 1998.
\bibitem{santos-diversifying} R.~L.~T.~Santos, C.~Macdonald, I.~Ounis. Selectively Diversifying Web Search Results. CIKM, 2010.
\bibitem{santos-xquad} R.~L.~T.~Santos, C.~Macdonald, I.~Ounis. Exploiting Query Reformulations for Web Search Result Diversification. WWW, 2010.
\bibitem{santos-survey} R.~L.~T.~Santos, C.~Macdonald, I.~Ounis. Search Result Diversification. Foundations and Trends in Information Retrieval, 9(1), 2015.
\bibitem{agrawal-diversify} R.~Agrawal, S.~Gollapudi, A.~Halverson, S.~Ieong. Diversifying Search Results. WSDM, 2009.
\bibitem{clarke-novelty} C.~L.~A.~Clarke, M.~Kolla, G.~V.~Cormack, O.~Vechtomova, A.~Ashkan, S.~B\"uttcher, I.~MacKinnon. Novelty and Diversity in Information Retrieval Evaluation. SIGIR, 2008.
\bibitem{drosou-diversity-survey} M.~Drosou, E.~Pitoura. Search Result Diversification. SIGMOD Record, 39(1), 2010.

%% --- Submodular optimization for selection / summarization ---
\bibitem{lin-bilmes} H.~Lin and J.~Bilmes. A Class of Submodular Functions for Document Summarization. ACL-HLT, 2011.
\bibitem{nemhauser-submodular} G.~L.~Nemhauser, L.~A.~Wolsey, M.~L.~Fisher. An Analysis of Approximations for Maximizing Submodular Set Functions. Mathematical Programming, 14(1), 1978.
\bibitem{khuller-budget} S.~Khuller, A.~Moss, J.~S.~Naor. The Budgeted Maximum Coverage Problem. Information Processing Letters, 70(1), 1999.
\bibitem{krause-submodular} A.~Krause, D.~Golovin. Submodular Function Maximization. In Tractability: Practical Approaches to Hard Problems. Cambridge University Press, 2014.
\bibitem{celf} J.~Leskovec, A.~Krause, C.~Guestrin, C.~Faloutsos, J.~VanBriesen, N.~Glance. Cost-Effective Outbreak Detection in Networks. KDD, 2007.

%% --- LLM agent memory and architecture ---
\bibitem{memgpt} C.~Packer, V.~Fang, S.~G.~Patil, K.~Lin, S.~Wooders, J.~E.~Gonzalez. MemGPT: Towards LLMs as Operating Systems. arXiv:2310.08560, 2023.
\bibitem{mem0} Mem0: A Memory Layer for LLM Applications. \url{https://github.com/mem0ai/mem0}, 2024.
\bibitem{generative-agents} J.~S.~Park, J.~C.~O'Brien, C.~J.~Cai, M.~R.~Morris, P.~Liang, M.~S.~Bernstein. Generative Agents: Interactive Simulacra of Human Behavior. UIST, 2023.
\bibitem{voyager} G.~Wang, Y.~Xie, Y.~Jiang, A.~Mandlekar, C.~Xiao, Y.~Zhu, L.~Fan, A.~Anandkumar. Voyager: An Open-Ended Embodied Agent with Large Language Models. TMLR, 2024.
\bibitem{react} S.~Yao, J.~Zhao, D.~Yu, N.~Du, I.~Shafran, K.~Narasimhan, Y.~Cao. ReAct: Synergizing Reasoning and Acting in Language Models. ICLR, 2023.

%% --- Memory benchmarks and evaluation ---
\bibitem{longmemeval} D.~Wu, H.~Wang, W.~Yu, Y.~Zhang, K.-W.~Chang, D.~Yu. LongMemEval: Benchmarking Chat Assistants on Long-Term Interactive Memory. ICLR, 2025.
\bibitem{locomo} A.~Maharana, D.-H.~Lee, S.~Tulyakov, M.~Bansal, F.~Barbieri, Y.~Fang. Evaluating Very Long-Term Conversational Memory of LLM Agents. ACL, 2024.
\bibitem{memagentbench} Y.~Hu, Y.~Wang, J.~McAuley. Evaluating Memory in LLM Agents via Incremental Multi-Turn Interactions (MemoryAgentBench). ICLR, 2026. arXiv:2507.05257.
\bibitem{liu-lost-in-middle} N.~F.~Liu, K.~Lin, J.~Hewitt, A.~Paranjape, M.~Bevilacqua, F.~Petroni, P.~Liang. Lost in the Middle: How Language Models Use Long Contexts. TACL, 2024.

%% --- Prompt compression and context engineering ---
\bibitem{llmlingua} H.~Jiang, Q.~Wu, C.-Y.~Lin, Y.~Yang, L.~Qiu. LLMLingua: Compressing Prompts for Accelerated Inference of Large Language Models. EMNLP, 2023.
\bibitem{llmlingua2} Z.~Pan, Q.~Wu, H.~Jiang, M.~Xia, X.~Luo, J.~Zhang, Q.~Lin, V.~R\"uhle, Y.~Yang, C.-Y.~Lin, H.~V.~Zhao, L.~Qiu, D.~Zhang. LLMLingua-2: Data Distillation for Efficient and Faithful Task-Agnostic Prompt Compression. ACL Findings, 2024.
\bibitem{recomp} F.~Xu, W.~Shi, E.~Choi. RECOMP: Improving Retrieval-Augmented LMs with Compression and Selective Augmentation. ICLR, 2024.

%% --- Tooling and systems ---
\bibitem{langchain} H.~Chase. LangChain. \url{https://github.com/langchain-ai/langchain}, 2022.
\bibitem{llamaindex} J.~Liu. LlamaIndex. \url{https://github.com/run-llama/llama_index}, 2022.
\bibitem{faiss} M.~Douze, A.~Guzhva, C.~Deng, J.~Johnson, G.~Szilvasy, P.-E.~Mazar\'e, M.~Lomeli, L.~Hosseini, H.~J\'egou. The Faiss Library. arXiv:2401.08281, 2024.

\end{thebibliography}
\end{document}